\documentclass[manuscript]{aastex}

\shorttitle{Central Star of EGB 6}
\shortauthors{Liebert, Bond, et al.}

\slugcomment{\bf ApJ, in press }

\begin{document}

\title{{\it Hubble Space Telescope\/} Imaging of the Binary Nucleus of
the Planetary Nebula EGB~6\altaffilmark{1}}

\altaffiltext{1} {Based on observations made with the NASA/ESA {\it
Hubble Space Telescope (HST)}, obtained by the Space Telescope
Science Institute. STScI is operated by the Association of
Universities for Research in Astronomy, Inc., under NASA contract
NAS5-26555. }

\author{James Liebert\altaffilmark{2}, Howard E. Bond\altaffilmark{3},
P.  Dufour\altaffilmark{4}, Robin Ciardullo\altaffilmark{5}, Michael
G.  Meakes\altaffilmark{6}, Alvio Renzini\altaffilmark{7}, and  
A. Gianninas\altaffilmark{8} }

\altaffiltext{2}{Steward Observatory, University of Arizona, Tucson AZ
85721; jamesliebert@gmail.com}

\altaffiltext{3}{Department of Astronomy \& Astrophysics, Pennsylvania
State University, University Park, PA 16802; Space Telescope Science
Institute, 3700 San Martin Drive, Baltimore, MD 21218; Current
address: 9615 Labrador Ln., Cockeysville, MD 21030; bond@stsci.edu}

\altaffiltext{4}{D\'epartement de Physique, Universit\'e de
Montr\'eal, C.P.  6128 Succ.\ Centre-Ville, Montr\'eal, Qu\'ebec
H3C~3J7, Canada; dufourpa@astro.umontreal.ca }

\altaffiltext{5}{Department of Astronomy \& Astrophysics, Pennsylvania
State University, University Park, PA 16802; rbc@astro.psu.edu}

\altaffiltext{6}{Space Telescope Science Institute, 3700 San Martin
Dr., Baltimore MD 21218; current address: 1717 Wilson Point Rd.,
Middle River, MD 21220; mgmeakes@gmail.com}

\altaffiltext{7}{Osservatorio Astronomico di Padova, Vicolo
dell'Osservatorio 5, I-35122, Padova, Italy;
alvio.renzini@oapd.inaf.it}

\altaffiltext{8}{Homer L. Dodge Department of Physics and Astronomy, 
University of Oklahoma, 440 W. Brooks St., Norman, OK 73019; 
alexg@nhn.ou.edu}

\begin{abstract}

EGB~6 is an ancient, low-surface-brightness planetary nebula.  The
central star, also cataloged as PG~0950+139, is a very hot DAOZ white
dwarf with an apparent M dwarf companion, unresolved from the ground
but detected initially through excesses in the {\it JHK\/} bands.  Its
kinematics indicate membership in the Galactic disk population. Inside
of EGB~6 is an extremely dense emission knot---completely unexpected
since significant mass loss from the white dwarf should have ceased
$\sim$$10^5$~yr ago.  The electron density of the compact nebula is
very high ($2.2\times10^6$~cm$^{-3}$), as indicated by collisional
de-excitation of forbidden emission lines.  {\it Hubble Space
Telescope\/} imaging and grism spectroscopy are reported here. These
resolve the white dwarf and apparent dM companion---at a separation of
$0\farcs166$, or a projected $96_{-45}^{+204}$~AU at the estimated distance
of $576_{-271}^{+1224}$~pc (using the $V$ magnitude).  Much to our
surprise, we found that the compact emission nebula is superposed on
the dM companion, far from the photoionizing radiation of the white dwarf.
Moreover, a striking mid-infrared excess has recently been reported in
{\it Spitzer\/}/IRAC and MIPS bands, best fit with two dust shells.
The derived ratio $L_{\rm IR}/L_{\rm WD} = 2.7\times10^{-4}$ is the
largest yet found for any white dwarf or planetary nucleus. The
compact nebula has maintained its high density for over three
decades. We discuss two possible explanations for the origin and
confinement of the compact nebula, neither of which is completely 
satisfactory. This leaves the genesis and confinement of the compact 
nebula an astrophysical puzzle, yet similar examples appear in the 
literature.

\end{abstract}


\section{Introduction: The Bizarre Nucleus of EGB~6}

\subsection{A Compact Nebula around the Central Star}

EGB~6 (PN~G221.6+46.4, J2000: 09 52 58.99, +13 44 34.9) is a large
($13'\times11'$) and very low-surface-brightness planetary nebula
(PN)\null.  It was serendipitously discovered by one of us (H.E.B.) in
1978, during examination of Palomar Observatory Sky Survey (POSS)
prints. The object was included in a list of faint nebulae found on
POSS prints published by Ellis, Grayson, \& Bond (1984, hereafter
EGB). An extremely blue, 16th-mag planetary nebula central star (CSPN)
is visible near the center of EGB~6. As described by EGB, a
spectroscopic observation revealed the surprising result that the
nucleus has strong [\ion{O}{3}] emission lines.  The central star was
detected independently as a faint high-latitude blue star in the
Palomar Green Survey (Green, Schmidt, \& Liebert 1986), designated
PG~0950+139; this was found to be a very hot DA white dwarf (WD) by
Fleming, Liebert, \& Green (1986). Spectroscopic observations by both
groups of authors found that the emission lines from the CSPN arise
from an unresolved compact emission nebula (CEN) component---a
startling result, given that the large post-asymptotic-giant-branch
age of the WD and EGB~6 would imply that significant mass loss from
the nucleus should have ceased long ago ($\sim$10$^5$ years). Note
that this approximate age is appropriate to the oldest known PNe, and
results from matching the slow kinematic velocity ($\sim$10~km/sec)
with the large size (a few parsecs) of the oldest PNe.  The CEN, in
addition to being confined to the central part of EGB~6, is far too
strong to be associated with the extremely faint outer PN, which to
our knowledge has not been observed spectroscopically. Deep CCD images
of EGB~6 have been published by Jacoby \& van de Steene (1995) and
Tweedy \& Kwitter (1996).

In a comprehensive study of the CEN by Liebert et al.\ (1989,
hereafter L89), the mystery intensified with the development of
quantitative results.  The very hot WD/CSPN is of type DAOZ, with a
hydrogen-rich atmospheric composition, but with \ion{He}{2}
4686~\AA\ appearing in absorption in the optical spectrum. Lines of
heavier elements were found in the ultraviolet by Gianninas et
al.\ (2010, hereafter G10), who published {\it FUSE\/} far-ultraviolet
spectra revealing ions such as \ion{C}{4}, \ion{N}{4}, \ion{O}{6},
\ion{P}{5}, \ion{S}{6}, \ion{Si}{4}, and \ion{Fe}{6}--{\small VII} in
the spectrum.

In a study of hot WDs from the PG Survey by Liebert, Bergeron, \&
Holberg (2005), the derived parameters for the central star of EGB~6 
were $T_{\rm eff} = 108,390\pm16,687$~K and $\log g = 7.39\pm0.38$.  
However, the model atmospheres used in that study assumed a pure hydrogen
composition and local thermodynamic equilibrium (LTE)\null. Since helium
and heavier metals appear in the spectrum, and non-LTE effects are
almost certainly important at $T_{\rm eff}\simeq100,000$~K, this analysis
needed to be redone.  Taking these into account has always caused
$T_{\rm eff}$ to be decreased.  Sure enough, G10 fit the spectrum with
non-LTE atmospheres, determining $T_{\rm eff}=93,233$~K, $\log g = 7.37$,
and $\log\,\rm(He/H)= -1.60$, assuming also the presence of C, N, and O
at solar abundances. (No attempt was made to determine actual
abundances of elements heavier than helium in their study.)  G10
determined an estimated absolute magnitude $M_V=+7.0$ (see their
Table~2).  We revisit the absolute magnitude, and estimate the
distance in \S3.3, showing that both are quite uncertain.

The WD has a measured proper motion in the UCAC4 catalog 
(Zacharias et al.\ 2012) of
$\mu_{\alpha} = 0\farcs0148$~yr$^{-1}$, $\mu_{\delta} =
0\farcs0023$~yr$^{-1}$.  At the rough estimated distance of 576~pc,
this object has a tangential velocity of about 40~km~s$^{-1}$.  L89
reported radial-velocity measurements from MMT echelle spectra of
$+6.2\pm1.8$ (for \ion{He}{2} 4686~\AA) and $+19.2\pm9.4$~km~s$^{-1}$ 
(for H$\alpha$).  Using a weighted mean of +6.7~km~s$^{-1}$ for the 
radialvelocity, we estimate a Galactic space motion relative to the 
local standard of rest of $U,V,W = +28.2, +11.8, -14.8$ km~s$^{-1}$.  
At the likely distance of $\sim$400~pc above the Galactic plane (at 
Galactic latitude $46.4^\circ$), this is an old Galactic disk object, 
as concluded by L89.

L89 showed that the CEN has the nearly unique property that the
[\ion{O}{3}] lines are partially quenched, and the expected
[\ion{O}{2}], [\ion{N}{2}], and [\ion{S}{2}] forbidden lines are
totally absent, due to collisional de-excitation at the very high
density.  This requires an electron density in excess of $10^6\rm\,
cm^{-3}$, making the CEN near the center of the old PN EGB~6 one of
the densest known ionized hydrogen nebulae.  However, as discussed
later, another compact and apparently similar nebula---Tol~26, studied
by Hawley (1981)---is even denser.  Dense compact nebulae can also
occur in the early post-asymptotic giant phase and in symbiotic stars.

The parameters of the EGB~6 CEN were estimated in more detail, using a
photoionization model, by Dopita \& Liebert (1989, hereafter DL). The
CEN was found to have a mean density of $n_e =
2.2\times10^6$~cm$^{-3}$ and a very small filling factor of
$\log~f\simeq-3.44$.  From the observed line fluxes, DL concluded
that the CEN is ionization-bounded, with an ionized mass of only
$7\times10^{-10}\,M_{\odot}$.  The radius of the Str\"omgren sphere
was found to be only 38~AU ($1.23\times10^{10}$~km), assuming that it
surrounded the WD\null. The line widths from echelle spectra covering
H$\alpha$ and \ion{He}{2} 4686~\AA\ showed single, narrow-line
components, implying an upper limit to any expansion rate of the
nebula of 50 km~s$^{-1}$.

DL speculated (several years before the first discoveries of ``hot
Jupiters'' around nearby stars) that the CEN around the nucleus of
EGB~6 might arise by ablation of a Jovian planet located within a few
AU from the WD\null.  While the hypothetical planet would have a radius
similar to a low-mass dwarf star, its escape velocity at a Jupiter
mass would be an order of magnitude lower, and of the same order of
magnitude as the thermal velocities of photoionized material on the
hot side facing the WD. However, this idea turned out to be incorrect.

\subsection{A Cool Companion of the Hot Nucleus}

Near-IR (NIR) photometry of EGB~6 by Zuckerman, Becklin, \& McLean
(1991) detected an appreciable IR excess in the $K$ band over the
Rayleigh-Jeans tail of the WD---with a coarse measurement of $K =
15.3\pm0.2$---which they attributed to an M dwarf stellar companion.
Fulbright \& Liebert (1993) confirmed the excess NIR fluxes, finding
$J = 16.43\pm0.20$, $H = 16.08\pm0.09$, and $K = 15.63\pm0.04$
(mean of multiple observations).  These may be compared to the
calculated magnitudes of the WD of $J = 17.60\pm0.62$, $H =
16.78\pm0.19$, and $K = 15.98\pm0.07$.  These authors searched for
variability over an interval of 30~days, in order to test the
possibility that, instead of a normal M dwarf companion, the WD had a
substellar or Jovian-mass companion so close that it was being heated
on its facing side by ultraviolet radiation from the WD.  In this
case, the expected orbital period would be hours to at most a few
days.  However, no variability was detected.

Note that Wachter et al.\ (2003) reported $J = 16.43\pm0.114$, $H =
15.555\pm0.144$, and $K_s = 15.350\pm0.151$ from the Two
Micron All Sky Survey (2MASS)\null.  The 2MASS measures agree well
at $J$, but poorly at $H$ and $K_s$. The 2MASS $K_s$ band has a
shorter long-wavelength cutoff than the standard Johnson $K$~band.
Moreover, the detection of variability at mid-IR (MIR) wavelengths
(see the next subsection; Su et al.\ 2011) could imply variability
also at the $K$ and possibly the $H$ band.  However, the variability 
in $H$ of 0.525~mag is greater than the difference of 0.28 between the 
mean $K$ of Fulbright \& Liebert (1993) and the 2MASS $K_s$ value. 
The latter would arguably be less enhanced by an increased MIR excess. 
So we are left with no clear explanation of the changes in $H$ and 
$K$. 

It thus appeared likely that the WD has an ordinary M dwarf companion,
although to our knowledge no IR spectrum has been obtained to confirm
this.  Using the accurate $K$-band measurement, at an estimated
distance assumed at that time to be 450~pc, Fulbright \& Liebert
(1993) inferred an absolute $M_K = +7.7\pm0.7$, suggestive of a
companion of spectral type dM3.5--dM5.  At the larger but uncertain
$576_{-271}^{+1224}$~pc estimate (using the $V$ magnitude), it is difficult
to estimate the spectral type of the apparent dM companion.  Moreover,
the $K$ magnitude is apparently variable.  Using relations from Veeder
(1974) for low-velocity main-sequence stars, we do note that a
spectral type of M4~V would correspond to an $M_{\rm bol} = +9.2$.

\subsection{Dust Components}

More recently, EGB~6 has been observed with {\it Spitzer\/} by Chu et
al.\ (2011). They found striking MIR excesses, in all four IRAC bands,
and with MIPS at 24~$\mu$m, at levels nearly four orders of magnitude
greater than the extrapolated WD photosphere---\S~4.  Bil\'ikov\'a et
al.\ (2011, 2012) showed that the infrared excess extends beyond
35~$\mu$m, and has temporal variations on a timescale of a year.  The
contribution of the M dwarf at these MIR wavelengths is expected to be
minor.

The MIR excess was best modeled with two dust shells of 500~K and
150~K, with a combined $L_{\rm MIR}/L_{\rm WD} = 2.7\times10^{-4}$---Chu
et al.\ (2011), excluding the contribution of the M dwarf.  This is the
largest IR excess found by these authors in a survey of many PN nuclei
and hot WDs.  Assuming an estimated distance of 647~pc, they
calculated $L_{\rm IR}/L_{\rm WD} = 0.026$ and 0.025~$L_{\odot}$, or 
$L_{\rm IR}/L_{\rm WD} = 2.68\times10^{-4}$ and $2.60\times10^{-5}$ for 
the cool and warmer shells, respectively (Bil\'ikov\'a et al.\ 2012).  
The emitting areas of the shells are 15 and 0.12 AU$^2$, respectively.

Even more remarkably, Su et al.\ (2011) discuss variations in this MIR
excess on a timescale of a year.  A followup observation with the {\it
  IRS} spectrograph shows that the dust continuum is detected out to a
wavelength of 35$\mu$m (Chu et al.\ 2011).  As noted earlier, it is
likely that the flux at $K$ and $H$ contribute to the high-frequency
tail of this MIR component---see Figure~2 of Chu et al.\ (2011).

Su et al.\ (2011) also discuss an observation of the CEN of EGB~6
obtained with the echelle spectrograph on the Kitt Peak 4-m Mayall
telescope.  The partially quenched [\ion{O}{3}] 5007~\AA\ profile
shows a double-peaked component, although the H$\alpha$ and H$\beta$
lines show only single peaks. The weaker [\ion{O}{3}] 4959~\AA\ 
probably also shows a doubled structure, and it is unclear what to 
conclude from the weak \ion{He}{1} 5876~\AA\ profile.  The full widths 
at half maximum (FWHMs) of all lines measured in the echelle spectrum 
appear to be basically the same.  The L89 MMT echelle observations of 
H$\alpha$ and \ion{He}{2} 4686~\AA\ were also single-peaked, with 
similar FWHM to the 4-m echelle values.  The doubling of the forbidden 
oxygen transitions may be a radiative-transfer effect, indicating that 
the collisional de-excitation is greatest near the line center. If 
instead the [\ion{O}{3}] profile were showing two velocity peaks, the 
same effect would presumably have been seen in the permitted lines.  
This proposed explanation needs to be tested by radiative transfer 
calculations also addressing the role of thermal broadening.

It was our hope in the early 1990s that additional clues to solving
the mysteries of the CEN component might be obtained from
high-resolution imaging with the {\it HST\/}, even with its compromised
capabilities at that time due to spherical aberration.  In \S~2 our
imaging results are presented, and we show that the CEN is detected as
a resolved point-source companion of the WD that coincides with the
apparent position of the dM star.  In \S~3 recent optical spectra
spanning another two decades are reported, and equivalent widths of
emission lines from the CEN are compared with prior values. In \S~4
the puzzling {\it Spitzer} infrared observations are reviewed.  In
\S~5 we attempt the daunting task of explaining the observations.
Conference papers summarizing some preliminary conclusions of our
study have appeared in Bond et al.\ (1992, 1993, 2009), Bond (1993),
and Liebert et al.\ (2013).

\section{Imaging Observations with HST}

We obtained five sets of {\it HST\/} images and grism spectroscopy
between 1991 and 1995, as summarized in the observing log in
Table~1. The first sequence of observations was obtained in 1991,
using the aberrated telescope with the Faint Object Camera (FOC) in
its f/96 mode. The narrow-band F486N and F501N filters were used, to
isolate the emission lines of H$\beta$ and [\ion{O}{3}] 5007~\AA,
respectively.  At the time of these observations, our aim had been to
test the Jovian-planet hypothesis of DL, in which case the CEN would
essentially coincide with the hot central star.  To our surprise, we 
instead resolved the system into two separate star-like objects, as 
depicted in the top row of images in Figure~1. The fainter component, 
brighter in [\ion{O}{3}] 5007~\AA, lies almost directly west of the 
primary star at separation measured initially to be $0\farcs173$.

We used the ``imexamine'' task in IRAF\footnote{IRAF is distributed by
the National Optical Astronomy Observatories, which are operated by
the Association of Universities for Research in Astronomy, Inc.,
under cooperative agreement with the National Science Foundation.}
to perform simple aperture photometry in order to measure the
magnitude differences between the two components; the results, which
are only approximate because the companion lies in the wings of the
aberrated images of the primary, are 2.3 and 1.3~mag, in H$\beta$ and
[\ion{O}{3}] respectively.  We may compare these to the predicted
magnitude differences between these two emission lines and the
continuum of the hot WD, as follows. The equivalent widths of H$\beta$
and [\ion{O}{3}], given in Table~2C of L89, are 2.9 and 17.9~\AA,
respectively. The FWHM widths of the F486N and F501N filter bandpasses
are 34 and 74~\AA\ (Nota et al.\ 1996). Thus the expected magnitude
differences are roughly $-2.5\log(2.9/34) = 2.7$ and
$-2.5\log(17.9/74) = 1.5$.  The rough agreement with the observations
shows that, at this level of approximation, essentially all of the
emission-line fluxes are located with the apparent companion dM
object, and not from the hot WD!

In 1993 February, in order to explore the relationship of the resolved
CEN source to the putative dM companion star, we obtained broad-band
images of EGB~6 with the original aberrated Wide Field and Planetary
Camera (WF/PC1) on {\it HST\/} in the high-resolution Planetary Camera
mode.  We used the F555W (``$V$'') and F785LP (``$I$'') filters. The
second row of pictures in Figure~1 depicts these images.  In a
separate ``Snapshot'' program (described in detail by Ciardullo et
al.\ 1999), in 1993 December, we observed EGB~6 again with WF/PC1 in
the same filters; these images are very similar to those illustrated
in Figure~1 and are not shown separately.  In these broad-band images,
we clearly detect a stellar companion, at the same separation and
position angle as the emission-line source. As the middle row in
Figure~1 shows, the companion is very red, being much brighter in the
$I$ band than in $V$. We conclude that this red source is the dM
companion that is the source of the NIR excess, and that the CEN is
associated with this object rather than the nearby hot WD. The source
is also weakly detected in F555W, but this filter transmits both
H$\beta$ and the [\ion{O}{3}] emission lines, which plausibly account
for the signal at $V$.

The magnitude difference between the $I$-band companion and the WD is
approximately 1.2~mag, but this is uncertain because the companion
unfortunately lies on the first Airy diffraction ring of the WD
primary. Roughly speaking, due to the large uncertainty in distance,
the red companion has $M_I\simeq+8.4\pm1.6$, in approximate
agreement with the conclusions about the red companion outlined in
\S~1.2, based independently on the {\it JHK\/} excess.

In order to verify the remarkable finding that the CEN is entirely
separated from the hot WD, in 1993 April we made low-dispersion
slitless spectroscopic observations with WF/PC1 in its grism
mode. Dispersed images were obtained with the G450L (covering the
interval 3000--6000~\AA\ at a dispersion of $25\,\rm\AA\,pix^{-1}$)
and G800L (6000--10000~\AA, $51\,\rm\AA\,pix^{-1}$) grisms.  We
selected a telescope roll angle that placed the dispersion
perpendicular to the line joining the two point-like sources, thus
maximizing their spatial separation.  Figure~2 is a pictorial
representation of the grism spectra.  In the G450L image, the
companion is detected only in the [\ion{O}{3}] 4959 and
5007~\AA\ emission lines (Figure~2 left panel; the much weaker
H$\beta$ line is not detected in these exposures). The emission
features are separated from the bright continuum of the WD by the same
angular amount as in the direct images.  In the G800L frame (Figure~2
right panel) we see flux from the companion only at the H$\alpha$
emission line.  At first glance, the companion appears also to have a
``pseudo-continuum'' in both images; however, these apparent continua
are artifacts produced by the Airy ring of the bright WD, as
demonstrated by the fact that their separation from the WD increases
proportionally to wavelength, as well as the fact that there are
identical artifacts symmetrically placed on the east side of the WD.

There are no detectable emission lines in the spectrum of the
hot WD or its immediate environment. As a further test, we determined 
a very approximate equivalent width (EW) for the combined flux from
[\ion{O}{3}] 4959 and 5007~\AA\ relative to the adjacent WD
continuum. This measurement is very crude because of the superposed
Airy ring, but subtracting it approximately from the line flux yields
a total EW of 18~\AA\null. This is in quite reasonable agreement, given 
the uncertainties, with the total of 23.3~\AA\ measured from calibrated
slit spectra of the combined light of both components (Table~2C in
L89). We conclude that the entire CEN flux comes from the immediate
vicinity of the dM companion, and not from the hot central star.

Our final {\it HST\/} imaging of EGB~6 was done in 1995 October, in a
continuation of our snapshot program on planetary nuclei. These
observations were done with the non-aberrated Wide Field Planetary
Camera~2 (WFPC2) and F555W (``$V$'') and F814W (``$I$'')
filters. Pictorial representations of these images are shown in the
bottom row of Figure~1. The faint companion of the CSPN is weakly
detected in both images. As before, the $V$ detection is most likely
due to the fact that the F555W filter transmits the H$\beta$ and
[\ion{O}{3}] emission lines. The red companion is detected in the $I$
band, but more weakly than in the earlier WF/PC1 F785LP image shown in
Figure~1. This apparent discrepancy is likely due to the fact that the
WFPC2 F814W band pass cuts off sharply near 9600~\AA, whereas the
WF/PC1 F785LP bandpass transmits up to about 1.1~$\mu$m. Examination
of Fig.~3 of Fulbright and Liebert (1993) suggests that the excess of
the dM companion over the declining WD continuum turns on sharply near
or slightly below 1~$\mu$m---see also Fig.~2 of Chu et
al.\ (2011). Thus the stronger detection of the companion with the
WF/PC1 $I$ filter than with WFPC2 is consistent with its very red
color.

We carried out astrometry of the {\it HST\/} images, using the
centroiding tool in IRAF's ``imexamine'' task, and conversion to right
ascension and declination with the ``xy2rd'' task. The results are
given in Table~2. The companion was not detected well enough in the
1993 December snapshot frames (which were single exposures, rather
than the pairs of frames taken at the other WF/PC1 and WFPC2 epochs
for cosmic-ray removal), so this epoch is omitted. The internal errors
of the astrometry are of the order of a few milliarcseconds, but it is
difficult to estimate the systematic errors due to the aberrated
images at the first two epochs, and the fact that the companion lies
in the Airy ring of the primary star.  The errors listed in Table~2
were estimated from the scatter among the three independent
measurements (thus tacitly assuming no significant orbital motion of
the companion from 1991 to 1995---a reasonable assumption for the
nominal orbital period estimated below---as well as assuming that the
emission-line object coincides exactly with the dM star).

The mean of the measurements (giving double weight to the FOC data) is
a separation of $0\farcs166$ and a J2000 position angle of
$267\fdg8$. At the estimated distance of $576_{-271}^{+1224}$~pc, this
corresponds to a projected separation of $96_{-45}^{+204}$~AU, and a
nominal orbital period (for assumed masses of 0.6 and $0.2\,M_\odot$) of
about 1200~yr. The linear dimensions of the surrounding faint, large
PN (EGB~6) are approximately 2--2.5~pc.

\section{Optical Spectra of EGB~6 Obtained Since 1988}

The available spectroscopic observations of EGB~6 from 1978 through
1987 were presented by L89. In this section we discuss two
observations obtained by us in subsequent years. The main purpose was
to extend the time baseline in a search for spectrum variability of
the CEN that might arise from changes in nebular density.

\subsection{1992 MMT Spectrum}

On 1992 December 27 we used the old Smithsonian-Arizona Multiple
Mirror Telescope (MMT, at that time with six 72-inch mirrors) and
red-channel spectrograph to observe EGB~6 with an $800\times1200$
Loral CCD detector and a $600\,\rm grooves\, mm^{-1}$ grating covering
3650--5250\AA\ at 6\AA\ resolution.  A long slit was used, but no
record of the slit width or positional orientation has survived for
this 20-year-old observation.  It was noted, however, that three
spectra were obtained under less-than-ideal weather conditions, and
were combined to achieve a continuum signal-to-noise (S/N) ratio near
80. Because the conditions were non-photometric, no absolute line
fluxes were determined. This spectrum is similar to prior published
spectra, except for the lower spectral resolution and higher S/N\null.
Unfortunately with no digital record, we cannot apply the method used
in the next subsection to subtract the photospheric continuum of the
modeled WD spectrum, published in 2010.

No significant variation in the relative strengths of the forbidden
lines compared to those reported in L89 was found.  In particular, the
ratio of the three [\ion{O}{3}] lines is very density-sensitive in the
high-density regime found in this object.  The ratio $R({\rm[O\,III]})
= [I({5007\,\rm\AA}) + I({4959\,\rm\AA)}]/I({4363\,\rm\AA})$ in the
1992 observation is identical to the value of 12.9 from the mean
values of the 1978--1987 observations from Table~2C of L89. Arguably
the most photometric measurement in L89 was the original IIDS
observation made by H.E.B. with the Kitt Peak 2.1-m telescope on 1978
April 11, using a $4\farcs3$ circular aperture; this observation gave
$R({\rm[O\,III]}) = 12.6$, not a significantly different value.  The
time span is 14~years between these measurements.

The [\ion{Ne}{3}] 3868.8~\AA\ line is not significantly different
(based on the EW) between the measurements of L89 (1.7~\AA) and 1992
(1.9~\AA). The critical density for this transition listed in
Osterbrock (1989, see Table 3.11) should also be low enough for
collisional de-excitation of [\ion{Ne}{3}] to occur at the DL nebular
density.  [\ion{Ne}{3}] 3868.8~\AA\ and 3967.5~\AA\ correspond to
[\ion{O}{3}] 4959 and 5007~\AA\, while the so-far unobserved
[\ion{Ne}{3}] 3342.5~\AA\ line corresponds to [\ion{O}{3}] 4363~\AA.
The neon ratios can similarly be measured as another test of the
electron density.

\subsection{2007 MMT Spectrum } 

The CEN presents a problem in determining the absolute strengths of
the Balmer and helium emission lines from ground-based spectra,
because they are superposed on photospheric absorption lines in the WD
spectrum. The best model-atmosphere fit to the WD spectrum is that of
Gianninas et al.\ (2010), which is the first non-LTE analysis
including CNO and helium abundances.  The model's synthetic spectrum
gives us a tool to remove the photospheric contamination of the CEN
spectrum.  A newer spectrum from the MMT, now a 6.5-m single-mirror
telescope, was obtained on 2007 December 15. The spectrograph, with a
$500\,\rm grooves\, mm^{-1}$ grating and one arc-second slit, achieved
4~\AA\ resolution.  However, the sky was partly cloudy, which stopped
us from observing at times during the night, so absolute fluxes could
not be measured. 

The top panel in Figure~3 shows this spectrum. The synthetic spectrum
from the best-fitting WD model of Gianninas et al.\ was then
subtracted, in order to obtain the CEN spectrum shown in the bottom
panel of Figure~3.  The subtraction of the model spectrum produced
clean nebular line fluxes, with no residual continuum, except for a
slight ``negative'' apparent flux in the 3900--3950~\AA\ region.

Table~3 presents the CEN fluxes, both the adopted values from L89
(corrected for a typographical error in the caption of the L89
Table~2C), and from the 2007 MMT observation. The fluxes are presented
in units of $10^{-15}\,\rm erg\,cm^{-2}\,s^{-1}$ for (1)~the quoted
mean L89 values (with no photospheric absorption corrections), and
(2)~those measured from the 2007 MMT observation, plotted in the
bottom panel of Figure~3, with correction for the WD absorption. Note
that the H$\alpha$ flux from the earlier measurements is affected
rather little by the weaker absorption line from the photosphere
because the H$\alpha$ emission is relatively strong.  The ratio of
H$\alpha$/H$\beta$ is 2.67 from the mean L89 flux values, but the
value of H$\beta$ is diminished by stronger photospheric absorption
than is H$\alpha$.  The ratio is close to the value of 2.81 which is
expected from an optically thick (Case~B), ionized hydrogen gas of
10,000~K, $n_e= 10^6\,\rm cm^{-3}$, the closest among the Tables
4.2--4.4 from Osterbrock (1974) to the $T_e$ and $n_e$ values
determined by DL.  The H$\gamma$/H$\beta$ and H$\delta$/H$\beta$ flux
ratios (to two significant digits) of 0.47 and 0.25 from the cleanest
emission spectrum of 2007 compare with 0.471 and 0.262 from this same
Osterbrock case.  Since H$\epsilon$ is blended with the even stronger
[\ion{Ne}{3}] 3967.5~\AA\ emission line, H$\epsilon$ must be omitted
from the discussion of ratios.  Finally, H8 (3889.1~\AA) and H9
(3835~\AA) also show the normal downward progression in strength
relative to H$\beta$. (Note that L89 did not comment on the higher
Balmer emission lines, presumably because of their weakness against
the WD photospheric absorption.)

Note that the actual $n_e$ is over a factor of two higher and
the $T_e$ of 11,400~K is over 10\% higher, but the dependence of the
ratios on $n_e$ and $T_e$ are modest.  The Balmer line ratios thus
indicate that very little reddening affects the CEN.

Some peculiarities are apparent in Table~3 and Figure~3. First, the
2007 fluxes are lower than those in L89 by an approximately constant
factor of $\sim$0.75. This likely indicates an error in the absolute
flux calibration of one (or both) of the observations, but the line
intensity ratios would remain reliable.

There is a broad apparent dip in absorption with a sharper dip
near 3930~\AA\ seen in Figure~3.  It is implausible for this to be
interstellar extinction, given the tight constraints on reddening
shown in L89 (their \S~2).  This apparent absorption dip is due to
imperfect subtraction of the WD synthetic spectrum.

The $R({\rm[O\,III]})$ value for the 2007 observation is 9.8,
apparently lower than the values of $\sim$12.9 in L89 and in
2002. Note, however, that the value of $R$ is very sensitive to the
weak [\ion{O}{3}] 4363~\AA\ line, which was probably effectively
weakened by a blend with the photospheric H$\gamma$ absorption line in
the earlier spectra, making the value of $R$ too high. There is thus
no strongly compelling evidence for significant changes in the CEN
spectrum over the 1978--2007 interval. Moreover, the equivalent widths
of the emission lines, i.e., normalized to the continuum of the WD and
independent of calibration zero-points, have not changed significantly
during this three-decade interval.

\subsection{Model-Atmosphere Fit}

We now revisit the G10 fit, which used the spectrum discussed in the
preceding subsection.  Figure~4 shows a more detailed display.  The
cores of the weak H and He~II absorption lines of the WD are
contaminated by the strong, narrow emission lines of the CEN\null.  The
detailed plot of individual line fits in the left panel of the figure
shows how difficult and inaccurate the fit is.  The surface gravity
with correct error bar is $\log g = 7.37\pm0.79$---see Fig.~23 of
G10.  The same figure shows that $T_{\rm eff}$ is also uncertain by
$\sim$20,425~K\footnote{The uncertainty in $T_{\rm eff}$ was mistakenly 
reported as 425~K in Gianninas, Bergeron, \& Ruiz (2011).}.  

These values were used to compute the absolute magnitude in the $V$ and
Sloan Digital Sky Survey (SDSS) $g$-bands, respectively, using the
photometric calibrations of Holberg \& Bergeron (2006).  These are
$M_V = +7.22\pm1.64$ and $M_g = +6.86\pm1.64$.  By coupling these
absolute magnitudes with the observed $V = 16.025\pm0.025$ (EGB) and 
$g = 15.66\pm0.04$, respectively, we computed spectroscopic distances 
of $576_{-271}^{+1224}$ and $585_{-276}^{+1243}$~pc. (We arbitrarily 
adopted the former value in some calculations herein.)  Though these 
distances are quite uncertain, they are at least consistent within 
the computed error bars.

\section{Constraints from the Mid-Infrared Excess } 

As discussed in \S~1.3, EGB~6 shows a MIR excess. The {\it Spitzer\/}
IRAC and MIPS fluxes of EGB~6, listed in Chu et al.\ (2011), at
wavelengths of 3.6~$\mu$m, 4.5~$\mu$m, 5.8~$\mu$m, 8.0~$\mu$m, and
24~$\mu$m are respectively $977\pm15$, $1176\pm15$, $1773\pm36$,
$3772\pm37$, and $11740\pm66$ $\mu$Jy.  These authors note
similarities to the MIR excess in the Helix PN, with an age of
$\sim$$10^4$~yr (O'Dell et al.\ 2002).  The {\it Spitzer\/} MIPS
observations of the Helix show a bright compact source in the 24 and
70~$\mu$m bands (Su et al.\ 2007).

For EGB~6, there is no indication of significant extinction in the
ultraviolet and optical fluxes of the hot star and the CEN\null.  For 
the star, L89 found that the {\it IUE\/} plus optical $UBV $ fluxes 
fit a power law of the following form ($F_\lambda$ in cgs units, 
$\lambda$ in \AA):

$$\log F_\lambda = -3.65\log\lambda - 1.175\, .$$ 

This was steeper than that of the well-studied 50,000~K DA WD
G~191-B2B (slope $-3.59$).  The conclusion was that there was no
detectable extinction, with a limit of $A(\rm1200\,\AA) < 0.2$~mag,
corresponding to $E(B-V) < 0.02$ (Seaton 1979).  This in turn
corresponds to $N_{\rm H} < 1.2\times10^{20}\,\rm cm^{-2}$, in the 
line of sight to the WD.  

For the CEN the extinction constraint is weaker, since the Balmer
emission line ratios are difficult to measure, being superposed on the
absorption lines of the WD\null.  Nonetheless, as noted earlier, L89 
found that the H$\alpha$/H$\beta$ flux ratio is consistent with no 
detected extinction, as are the ratios of higher Balmer lines 
discussed in \S~3.2.

For this line-of-sight position on the sky, Schlegel et al.\ (1998)
estimate $A_V = 0.104$, implying $E(B-V) = 0.034$; this results from 
a reprocessed composite map of the $COBE$/$DIRBE$ and $IRAS$/$ISSA$
maps, with zodiacal foreground and confirmed point sources removed.
Schlafly \& Finkbeiner (2011) measure $A_V = 0.086$, implying 
$E(B-V) = 0.028$, using measurements of dust reddening from the colors 
of stars with spectra obtained by the SDSS\null.  Here we used the 
tools provided by the NASA/IPAC Extragalactic Database (NED) for 
coordinate transformation and galactic extinction calculation.  In both 
cases a standard ratio, $R = A_V/E(B-V) = 3.1$, was applied to convert 
to $E(B-V)$.  Both estimates of $E(B-V)$ for this celestial position 
are again quite small, and apply to the total line of sight out of the
Galaxy.  Thus, there is no evidence that the source of the excess 
{\it Spitzer\/} MIR flux is in front of either the hot WD star or the 
CEN.

Assuming that the origin of the excess detected in the MIR is two
shells or disks of dust, as fitted by Bil{\'i}kov{\'a} et al.\ (2012),
we note that the surface areas---15~AU$^2$ for the 100~K shell, and
0.12~AU$^2$ for the 500~K shell---are large compared with the WD and
putative M dwarf.  Hence the column densities of dust in the 
respective lines of sight may not be large enough to register 
detectable extinction in either case.  This leaves undetermined the 
location of the presumed dust shells. 

\section{The Compact Emission Nebula (CEN)}

\subsection{Constraints on the Model}

As stated in \S2, the angular separation of $0\farcs166$ implies a
lower-limit physical separation of the CEN from the WD of
$a\simeq96_{-45}^{+204}$~AU for the assumed distance of 
$576_{-271}^{+1224}$~pc.  The DL filling factor, $f$, might now be 
interpreted instead as a covering factor---i.e., that the ionized gas 
sits in a large CEN coincident with the $I$-band and NIR dM component, 
intercepting a fraction $f$ of the WD's radiation.  If spherical, and 
adopting the DL value of $f=3.63\times10^{-4}$, the CEN would have a 
radius given by

$$R_{\rm CEN} \simeq  2\,a\,f^{1/2} \simeq 3.6\,\rm AU \, ,$$

\noindent assuming the CEN to be optically thin in H$\beta$, as the
fits to Case B of Osterbrock (1974) indicate.  However, correcting for
the distance now estimated to be $576_{-271}^{+1224}$~pc, the filling/ 
covering factor $f$ will be $(6.4\pm3)\times10^{-4}$.  Now the above 
equation produces $R_{\rm CEN} = (4.2\pm2)\times10^8\,\rm km = 
4.8_{-2.3}^{+10.3}$~AU.

Note that this supergiant size is necessary in order for the CEN to
intercept enough ionizing photons to produce the observed H$\beta$
flux (see L89).

However, since the CEN does not surround the WD, the simple
steady-state mass-loss hypothesis envisioned by L89 and DL cannot
apply. If in fact the excited gas were due to ongoing mass loss from
the dM star, the CEN should be growing in size and in intercepted
solid angle of radiation from the WD\null. This would cause the 
luminosity of the CEN to grow with time, with its average density 
decreasing.  If the CEN were formed from gas lost from the M dwarf, 
this would also lead to a time-varying nebula; the nebula would grow 
in size and decrease in density.  However, as described in \S~3, over 
three decades of observations have not revealed any changes in the
emission-line spectrum.

Instead of a slowly expanding gaseous nebula recombining at the
ionization boundary and replenished by mass loss from the host object,
we are forced to invoke a scenario involving an essentially static
CEN of extraordinarily high density---in comparison to standard
\ion{H}{2} regions and normal PNe (apart from the most compact, young
proto-planetary nebulae, and a few peculiar objects like those
discussed below in \S~6). Note, however, that the emission-line widths
are consistent with an expansion velocity of up to about 50
km~s$^{-1}$.  But, if from the nonvariability argument, we assume
simply that the CEN has not doubled in size in over twenty years,
an upper limit to the expansion velocity is $<$1~km~s$^{-1}$!

We now consider two hypothetical explanations for the existence of a
CEN surrounding, or in the immediate vicinity of, the dM companion of
the hot CSPN: (1)~it is a region of compressed gas, created at the
location where a fast wind from the WD collides with a slower wind
from the dM star; or (2)~it is a residual structure around the dM
star, consisting of material captured from the PN outflow from the
CSPN, and surviving to the present time.

\subsection{The ``Dueling-Winds'' Model}

Let us assume that both binary components are losing mass in the form
of winds.  As noted in the previous subsection, any such wind from
either component is expected to have a velocity far in excess of the
measured limit on the photo-ionized gas discussed in the previous
section.  Let us further hypothesize, however, that the faster WD wind
encounters that of the dM dwarf near the latter's position in the
binary system (as required by the imaging of the CEN component). Given
sufficient mass fluxes and supersonic velocities, a shock may result
which would greatly slow down both gaseous components.  The density of
the shocked gas can be much higher than the densities of either wind
component because it is confined by the ram pressures of the two
winds.  This essentially static entity will then be photoionized by
radiation from the WD.

The constraints on this model from the observations are:

(1). The ram pressures of the two winds are comparable to the pressure
in the ionized gas, which is known from the ionization model to be

$$P_g = 3\,n_{\rm H}\,kT \simeq 1\times10^{-5} \, \rm dynes\,cm^{-2} \, ,$$

\noindent (using $n_{\rm H} = 2.2\times10^6\,\rm cm^{-3}$ and $T=
11,400$~K).

(2). The projected separation of the binary is $a = 96_{-45}^{+204}$~AU, 
so the true separation is at least this much.  Since this value and the
systemic distance are uncertain, we shall scale to a binary separation
of 100~AU in the calculations below.

(3). The covering factor of the ionized gas is also determined by the
ionization model to be approximately $f\simeq 6.4\times10^{-4}$.

We can derive the product $\dot{M}_{\rm WD} V_{\rm WD}$ for the WD
from the first two constraints, and the same product ($\dot{M}_{\rm
 dM} V_{\rm dM}$) for the companion using the third.  For the WD
\begin{eqnarray*}
\dot{M}_{\rm WD} V_{\rm WD} &=& 4\pi a^2 P_g /\beta\\
                            &=& 6\times10^{25} \, {\rm g \, cm\, s^{-2}} \\
               &=& 1\times10^{-5}(a/100AU)^2~M_\odot \, {\rm yr^{-1} \, km\, 
s^{-1}} \, ,\\
\noalign{or}
\dot{M}_{\rm WD} &=& 1\times10^{-9} \, M_{\odot}\, {\rm yr^{-1}} \,
\left({10,000 \rm\, km\, s^{-1}} \over {V_{\rm WD}}\right) , \\
\end{eqnarray*}
where $\beta$ is the overpressure (approximately unity, or a bit
larger) in the stagnation region between the shocks.

While there is no physical justification for such a high mass-loss
rate from a hydrogen-rich WD, it probably cannot be entirely ruled out
by observation.  This model might be tested by high-resolution
spectrophotometry covering the far-ultraviolet resonance
lines. Ground-based echelle spectra covering \ion{He}{2} 4686~\AA\ and
H$\alpha$ show single-component lines with no evidence for outflow,
but these are not resonance lines.

To derive $\dot{M}_{\rm dM}\, V_{\rm dM}$ for the companion, we need
to know how far from the dM star the shock is located.  As noted in
\S5.1, the radius of the CEN is about $R_{\rm CEN} \simeq
2\,a\,f^{1/2} \simeq 4.8(a/100AU)\,\rm AU$.

If we make the plausible assumption that the dM wind shock occurs at a
distance $R_{\rm cl}$ from the star, then
\begin{eqnarray*}
\dot{M}_{\rm dM}\, V_{\rm dM} &=& 4 f \dot{M}_{\rm WD}\, V_{\rm WD} \\
&=& 1.4\times10^8\, M_{\odot} \, {\rm yr^{-1}\,km\,s^{-1}} \, ,\\
\noalign{yielding}
\dot{M}_{\rm dM} &=& 1.4\times10^{-11}\,M_{\odot}\,{\rm yr^{-1}} 
  \left({1000\, \rm km\,s^{-1}} \over {V_{\rm dM}}\right) \, . \\
\end{eqnarray*}

There are no recent estimates of the mass loss rate from M dwarfs,
after an old, highly uncertain figure of
$\sim$$10^{-12}\,M_\odot$yr$^{-1}$, ascribed to flaring M dwarfs
(Coleman \& Worden 1976), although Wood et al.\ (2001, 2002) present 
a much lower upper limit for the less active M5.5~V solar neighbor
Proxima Centauri.  Moreover, in the present case the existence of a
hot companion may enhance whatever spontaneous mass loss the M dwarf
may sustain.  Actually, we can turn around the issue, and claim that
ours is a reasonable estimate for the current mass loss rate from this
M dwarf companion to a hot white dwarf.  Measuring mass loss rates
from red giants is already a very uncertain exercise, while measuring
it in M dwarfs looks to be totally hopeless.

There is one further argument that can be used to restrict
$\dot{M}_{\rm dM}$.  The ionized gas must be flowing out of the
compressed region between the wind shocks at roughly the thermal
velocity, $v_{\rm ion} \simeq 10\,\rm km\, s^{-1}$.  Therefore the
timescale for the ionized gas outflow is
$$\tau_{\rm ion} \simeq  R_{\rm CEN}/v_{\rm ion} \simeq  2\,\rm yr\, .$$

Balancing the mass input from the dM star wind with the mass output of
the ionized gas, we have
$$\dot{M}_{\rm dM} M_{\rm ion}/\tau_{\rm ion}\simeq 5\times10^{-10} M_\odot\,
  \rm yr^{-1} \, ,$$
based on the observed mass of ionized gas, $M_{\rm ion} \simeq
7\times10^{-10}\,M_\odot$.  Note that the mass loss rate is appreciably 
larger than the estimate two paragraphs ago, but that estimate was 
judged to be uncertain. 

Finally, in \S~4 we showed that the extinction in the lines of sight
to the WD and CEN are small (undetectable). We assume that the CEN is
produced where the two winds collide between the WD and dM, but closer
to the latter.  Let us hypothesize that the dust condenses out behind
this nebula, where it is initially shielded from the hard radiation of
the WD\null.  However, as it spreads out, it will likely absorb WD flux 
and form a turbulent stream which drifts beyond the dM and CEN.  Note 
that the dust does not lie in front of (the line of sight to) either the 
WD or dM in this scenario.  It can then be heated by radiation from both
components, as the dust is slowly driven by radiation pressure away
from the WD\null.  Such a hypothesis might produce the only component of
this system that varies---the dust shell(s)---if the dust forms
episodically, then drifts away.

\subsection{A Residual Envelope or Circumstellar Disk around the M 
Dwarf}

An alternative explanation for the CEN in the EGB~6 system was
suggested by Zuckerman et al.\ (1991): that the dM star captured the
amount of nebular mass cited above into an envelope or circumstellar
disk, at the time of the PN ejection from the primary star.  It is
plausible to assume that a disk might form around the companion star,
since the orbital motion imparts a net angular momentum to the
outflowing gas (or it could also have a small amount of intrinsic
angular momentum).

Jeffries \& Stevens (1996) discuss a class of wide binaries containing
hot WDs or CSPN and cool, rapidly rotating, magnetically active
companions.  Significant spin-up of the latter may occur, resulting in
what they call a wind-induced rapidly rotating (``WIRRing'') star,
along with the formation of a disk, and accretion of chemically
enriched material from the AGB star.  The detection of peculiar carbon
and $s$-process elements such as barium that were present in the wind
of the former-AGB star may thus provide an evolutionary link between
WIRRing stars and barium giants.  This phenomenon may explain the
``Abell~35''-type PN nuclei, in which the hot CSPNs are accompanied by
rapidly rotating cool companions (e.g., Bond, Ciardullo, \& Meakes
1993).  Note, however, that the component separation in EGB~6 is much
larger than in the known barium binary stars, but this may simply be
due to selection against finding such systems with much wider
separations.

A good example may be WeBo~1 (PN~G135.6+01.0), having a remarkable
thin-ring morphology around a late-type giant with overabundances of
carbon and $s$-process elements (Bond, Pollacco, \& Webbink 2003). The
giant is chromospherically active, a ``spotted'' WIRRing star with a
4.7-day rotation period.  These authors surmised that the likely
undetected CSPN is now a hot subdwarf or WD which polluted its
companion, and spun up its rotation, during the PN ejection phase. The
putative hot companion of the cool optical barium star in WeBo~1 has
recently been detected directly through UV observations with the 
{\it  Swift\/} satellite (Siegel et al.\ 2012). Abell~70 is a another PN 
that may be a similar case, where a cool, optical, barium-enhanced
companion is detected along with the hot central star (Miszalski et
al.\ 2011,2012).  Mira~B is a WD $\sim$70~AU from the AGB primary, 
recently resolved by {\it HST\/} (Ireland et al.\ 2007). However, Mira~B 
apparently has an optically thick circumstellar disk regularly being fed 
material from the mass-loss/wind---see Fig.~3 of Ireland et al.\ (2007).

These examples make it clear that an accretion disk can form around a
companion to a PN nucleus (a former AGB star, now a WD), even at
fairly large separations.  However, the active PN-ejection phase
typically lasts no more than a few times $10^4$~years.  Whether a
dense disk or envelope around the late-type companion can survive for
$\geq$$10^5$~years is unclear.  This is the evolutionary time
estimated for the WD (L89), which is similar in order of magnitude to
the likely kinematic expansion age of the extended PN of EGB~6. 

Moreover, the required size of the presumed disk or CEN is again
constrained by the covering factor to be several AU\null.  This is
implausibly large, compared with the sizes of disks normally
associated with main-sequence stars.

If a captured circumstellar disk explains the CEN in EGB~6, then it is
puzzling why other PN nuclei in which a low-mass companion orbits the
CSPN at smaller separations than in EGB~6 do not show the same
phenomenon.  An example of such a system is BE~Ursae Majoris, a
short-period eclipsing binary located within a faint PN; this shows a
spectacular emission line spectrum, with strong orbital variations due
to the heating of the facing side of the companion (Ferguson et
al.\ 1987). However, there is no evidence for mass loss or the
formation of an ambient emission nebula.  A series of spectra of BE~UMa 
obtained during the 1990s (unpublished) both at minimum and near
maximum light show no evidence for [\ion{O}{3}] or [\ion{Ne}{3}], the
strong forbidden emission lines seen in the EGB~6 spectrum (Fulbright
\& Liebert 1993).

Finally, the origin of the dust shells must be in the outer part of
the disk in this scenario. This places it in the line of sight to the
secondary star, which might cause extinction, unless the dust torus
were at high inclination to our line-of-sight.  On balance, we favor
the scenario of the previous subsection, the ``duelling winds'' model.

\section{EGB~6: An Unsolved Problem of Stellar Evolution}

We end this paper in an unsatisfactory way: no clear answers, only the
accumulation of questions.  Our study of this object is now entering
its fourth decade, beginning with the discovery and first observations
in the 1970s, and the publications of the discoveries by EGB in 1984
and the PG Survey in 1986.  The mysteries seem only to compound. 

With the launch of {\it Spitzer}, new observational results were
introduced, which forced a reconsideration of the extant issues, but
pose at least two new ones.  What is the source of the MIR excess,
detected in all IRAC bands and with MIPS and IRS?  The simplest model
is to introduce two dust shells of 150 and 500~K\null.  But where is 
this dust located in or around the binary system?  Why is there 
absolutely no evidence of extinction in front of the WD, and (somewhat 
less established) in front of the CEN?  A residual shell complex on a 
much bigger scale applicable to a younger PN such as the Helix seems to 
be a possible explanation.  However, the EGB nebular shell is much 
older than the Helix, yet the CEN is very much smaller.  Perhaps the
scenario outlined in \S~5.1 is the best explanation.

It might be worthwhile to summarize the basic constraints on the 
model: 

1) there is no variability, except for the assumed MIR-emitting dust
shells.

2) the CEN is situated with the presumed dM companion star, which
dominates the $JHK$ flux, at least $\sim$96~AU from the source of 
photoionizing radiation.

3) neither the WD nor CEN are appreciably reddened. 

Point (1) implies a stationary configuration, with the CEN constantly
replenished to compensate for its evaporation, on a timescale of order
one year.  Point~(2) leads to an origin of the CEN in the colliding
winds in between the two stars, but closer to the dM\null. Point~(3)
suggests that the MIR emitting dust is in front of neither the WD nor
the CEN; hence its likely location is in a turbulent stream beyond the
dM and the CEN\null.  If the dust forms episodically, and later is 
lost from the system, this component could vary; this would account 
for the observed MIR flux variation.

Finally, we note that the ``EGB~6 phenomenon''---a very dense, compact
emission nebula (CEN) ---is not unique.  In a paper that pre-dated the
publication of the EGB~6 CEN, Hawley (1981) studied a ``peculiar
emission line object'' (1230$-$275, now designated Tol~26), found in a
Curtis Schmidt survey at Cerro Tololo by Smith et al.\ (1976).  Hawley
showed that this CEN is even denser than the one in EGB~6 ($n_e =
4\times10^6$~cm$^{-3}$), as indicated by the [\ion{O}{3}] line
ratios and the absence of the singly ionized forbidden O, N, and S
doublets.

Frew \& Parker (2010, their \S4.7.1, entitled ``EGB~6 and its Kin'')
discuss several other PNe and related objects with unresolved compact,
dense knots at their centers.  The central stars are not always WDs.  
For example, PHR~J1641$-$5302 has a nucleus classified WC4, a much 
more luminous star than in EGB~6.  However, the nebula shows an
$R(\rm[O~III])$ value indicating high density, again with no
[\ion{N}{2}] or [\ion{S}{2}] detected.  The nature of the very red
PHR~1757$-$1711 = PTB~15 (Boumis et al.\ 2003) is unclear, since
[\ion{O}{3}] 4363~\AA\ was not measured; however, the emission nebula
shows no [\ion{N}{2}] or [\ion{S}{2}].  NGC~6804 has a compact
emission nebula coincident with the luminous CSPN (Bil{\'i}kov{\'a} 
et al.\ 2012). Finally, Miszalski et al\. (2011) suggest that the 
EGB~6 nebula has evolved from an object like the PN M~2-29; this 
luminous object has shown an R~Coronae Borealis-like fading event in 
its light curve, triggering a dust/cloud formation event.

Thus, Nature has been able to reproduce what seem like very unusual
physical conditions at least a few times.

\section{Suggested Further Observations}

In lieu of having clear answers, we now mention some future work 
which can result in some clarification:

(1)~An ${\it HST}$/STIS spectrum of the resolved WD could 
produce much-improved stellar parameters, especially the surface 
gravity and absolute magnitude, and the spectrophotometric distance 
measurement.  A measurement of the distance by the European Space 
Agency {\it Gaia\/} mission may also be possible in a few years. 

(2)~An IR spectrum, even at low resolution, preferably using adaptive
optics, should identify and provide quantitative information on the
putative M dwarf in this system.

(3)~The most useful optical observations may be to continue to 
monitor the CEN spectrum to see whether evidence for a variation in 
nebular density occurs.

(4)~The variability of the MIR component needs to be investigated
further, hopefully by repeating observations with {\it Spitzer}.  A
spatially resolved MIR image may become possible with the launch of
the {\it James Webb Space Telescope}.  Certainly further study of 
the MIR spectrum will be possible with {\it JWST}. 

(5) The modelling of the nebula, as done over 20 years ago by DL, 
needs to be repeated with all the new observational data. 

\acknowledgments

Support for this work was provided by NASA through grant number
GO-2570 from the Space Telescope Science Institute, which is 
operated by AURA, Inc., under NASA contract NAS 5-26555. We thank 
Drs.\ John Bieging and Jay Holberg for help with logistics, and
Astronaut/Professor Steve Hawley for pointing out Tol~26.

{\it Facilities:} \facility{{\it Hubble Space Telescope} (WF/PC1,
WFPC2), MMT}

\newpage

\begin{deluxetable}{lllcll}
\tablenum{1}
\tablecaption{{\it HST} Imaging Observations of EGB~6} 
\tablewidth{0pt}
\tablehead{
\colhead{Date} &
\colhead{Camera} &
\colhead{Filter or} &
\colhead{Image Scale} &
\colhead{Exposure} &
\colhead{Program ID \& PI} \\
\colhead{} &
\colhead{} &
\colhead{Grating} &
\colhead{[$''$/pix]} &
\colhead{[s]} &
\colhead{} 
}
\startdata 
1991 Dec 2 & FOC {\it f}/96 & F486N (H$\beta$)     & 0.022 & 693 & GO-2570/Bond \\
1991 Dec 2 & FOC {\it f}/96 & F501N ([\ion{O}{3}]) & 0.022 & 796 & GO-2570/Bond \\
\noalign{\vskip0.05in}
1993 Feb 3 & WF/PC1/P6 &   F555W  ($V$) & 0.044 & $2\times100$ & GO-4778/Bond \\        
1993 Feb 3 & WF/PC1/P6 &   F785LP ($I$) & 0.044 & $2\times500$ & GO-4778/Bond \\  
\noalign{\vskip0.05in}
1993 Apr 20 & WF/PC1/P6 & G450L & 0.044 & $3\times1200$ & GO-4778/Bond \\  
1993 Apr 20 & WF/PC1/P6 & G800L & 0.044 & $3\times1800$ & GO-4778/Bond \\
\noalign{\vskip0.05in}
1993 Dec 1 & WF/PC1/P6 &   F555W  ($V$) & 0.044 & 200  & SNAP-4308/Bond \\      
1993 Dec 1 & WF/PC1/P6 &   F785LP ($I$) & 0.044 & 1000 & SNAP-4308/Bond \\ 
\noalign{\vskip0.05in}
1995 Oct 11 & WFPC2/PC &  F555W ($V$) & 0.046 &  80, 160 & SNAP-6119/Bond \\	 
1995 Oct 11 & WFPC2/PC &  F814W ($I$) & 0.046 & 350, 500 & SNAP-6119/Bond \\
\enddata 
\end{deluxetable}

\begin{deluxetable}{lcc}
\tablenum{2}
\tablecaption{Astrometry of the EGB~6 Companion Object} 
\tablewidth{0pt}
\tablehead{
\colhead{Date} &
\colhead{Separation [$''$]} &
\colhead{PA (J2000) [$^\circ$]} 
}
\startdata 
1991.9198 & $0.173\pm0.009 $ & $266.3\pm1.7$ \\
1993.0942 & $0.156\pm0.009 $ & $268.9\pm1.7$ \\
1995.7768 & $0.162\pm0.009 $ & $269.6\pm1.7$ \\
\enddata 
\end{deluxetable}

\begin{deluxetable}{lcccc}
\tablenum{3}
\tablecaption{EGB~6 Emission-Line Fluxes [$10^{-15}\,\rm erg\,cm^{-2}\,s^{-1}$]}
\tablewidth{0pt}
\tablehead{
\colhead{Ion} & 
\colhead{Wavelength [\AA]} & 
\colhead{Flux, L89} & 
\colhead{Flux, MMT 2007} & 
\colhead{Notes}  
}

\startdata 
\leavevmode [\ion{O}{2}] & 3727 & $<$1.2 & -- &1 \\ 
H9           & 3835 & -- & 0.2: & \\  
\leavevmode [\ion{Ne}{3}] & 3869 & 9.1 & 7.7 & \\
H8           & 3889 & -- & 0.8 & \\
H$\epsilon$+[\ion{Ne}{3}]  & 3970 & -- & 3.4 & \\
\leavevmode [\ion{S}{2}] & 4076 & $<$0.9 & -- &1 \\ 
H$\delta$    & 4101 & 1.7 & 1.2 & \\ 
H$\gamma$    & 4340 & 3.2 & 1.9 & \\
\leavevmode [\ion{O}{3}] & 4363 & 3.8 & 3.7 & \\
\ion{He}{1}  & 4471 & 0.6: & 0.4 & \\
\ion{He}{2}  & 4686 & 0.5: & 0.7 &2 \\
\leavevmode [\ion{Ar}{4}] & 4712 & 1.3 & ?? &3 \\
\leavevmode [\ion{Ar}{4}] & 4740 & 0.5: & ?? & \\ 
H$\beta$     & 4861 & 6.8 & 4.7 &2 \\
\leavevmode [\ion{O}{3}] & 4959 & 11.7 & 9.2 &1 \\
\leavevmode [\ion{O}{3}] & 5007 & 37.5 & 27.1 &1 \\
\leavevmode \ion{He}{1}  & 5876 & 1.1 & -- & \\ 
\leavevmode [\ion{O}{1}] & 6300 & 1.4 & -- & \\  
H$\alpha$    & 6563 & 18.2 & -- & \\ 
\ion{He}{1}  & 6678 & 0.5: & -- & \\ 
\leavevmode [\ion{S}{2}] & 6717,6731 & $<$0.7 & -- &1 \\ 
\leavevmode [\ion{O}{2}] & 7325 & 0.8: & -- & \\ 
\enddata

\tablecomments{
(1) measured flux partially or totally quenched due to high density;  
(2) measured flux in L89 are reduced by photospheric absorption and 
have not been corrected;  
(3) possible blend of [\ion{Ar}{4}] and \ion{He}{1}. 
} 

\end{deluxetable}

\begin{figure}
\begin{center}
\includegraphics[width=3.9in]{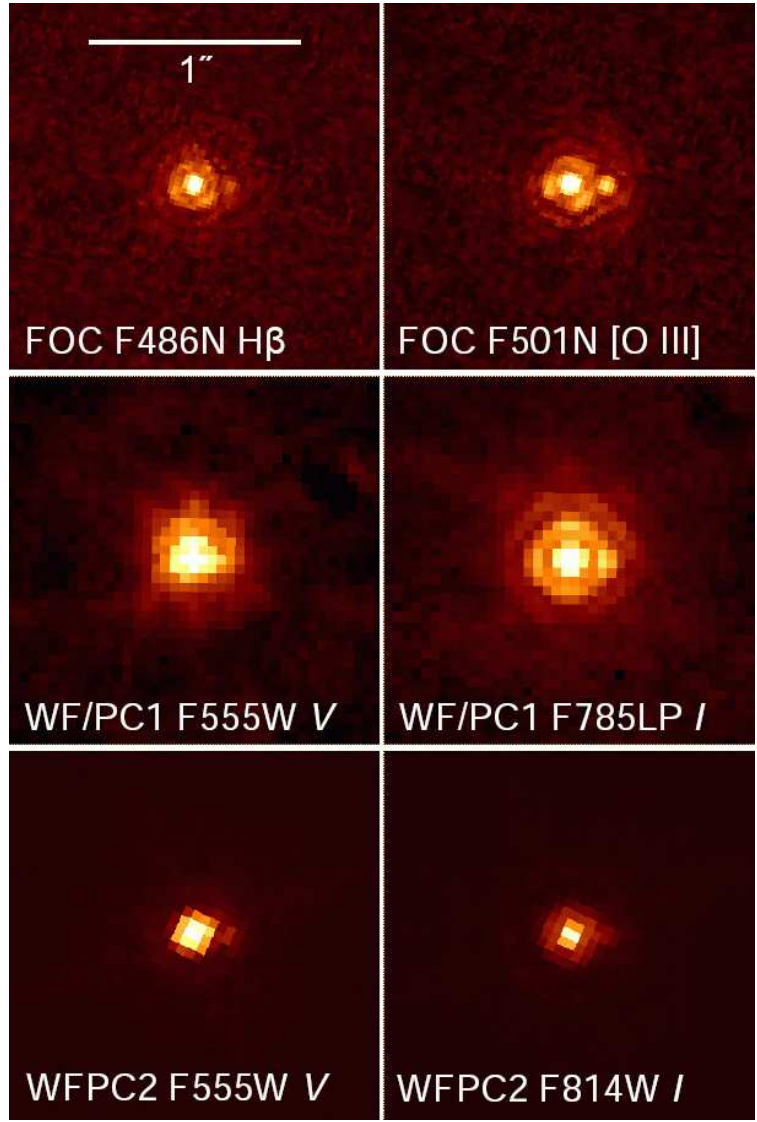}
\end{center}
\vskip-0.2in 
\caption{{\it HST\/} images of the central star of EGB~6.  North
is at the top and east on the left in all images, and each frame is
$1\farcs75\times1\farcs75$.  {\bf Top row:} Faint Object Camera
images taken in 1991 in narrow-band filters isolating H$\beta$ and
[\ion{O}{3}] 5007~\AA. The bright star is the hot white dwarf
central star, which is accompanied by an emission-line companion
located almost directly west. {\bf Middle row:} Wide Field/Planetary
Camera~1 images taken in 1993 in broad-band ``$V$'' and ``$I$''
filters. The cool dM companion is detected in the $I$ filter at the
location of the CEN. {\bf Bottom row:} Wide Field and Planetary
Camera~2 images taken in 1995 in broad-band ``$V$'' and ``$I$''
filters. The companion star is again detected, but is fainter in $I$
due to a narrower bandpass, as discussed in the text.}
\end{figure}

\begin{figure}
\begin{center}
\includegraphics[width=5.5in]{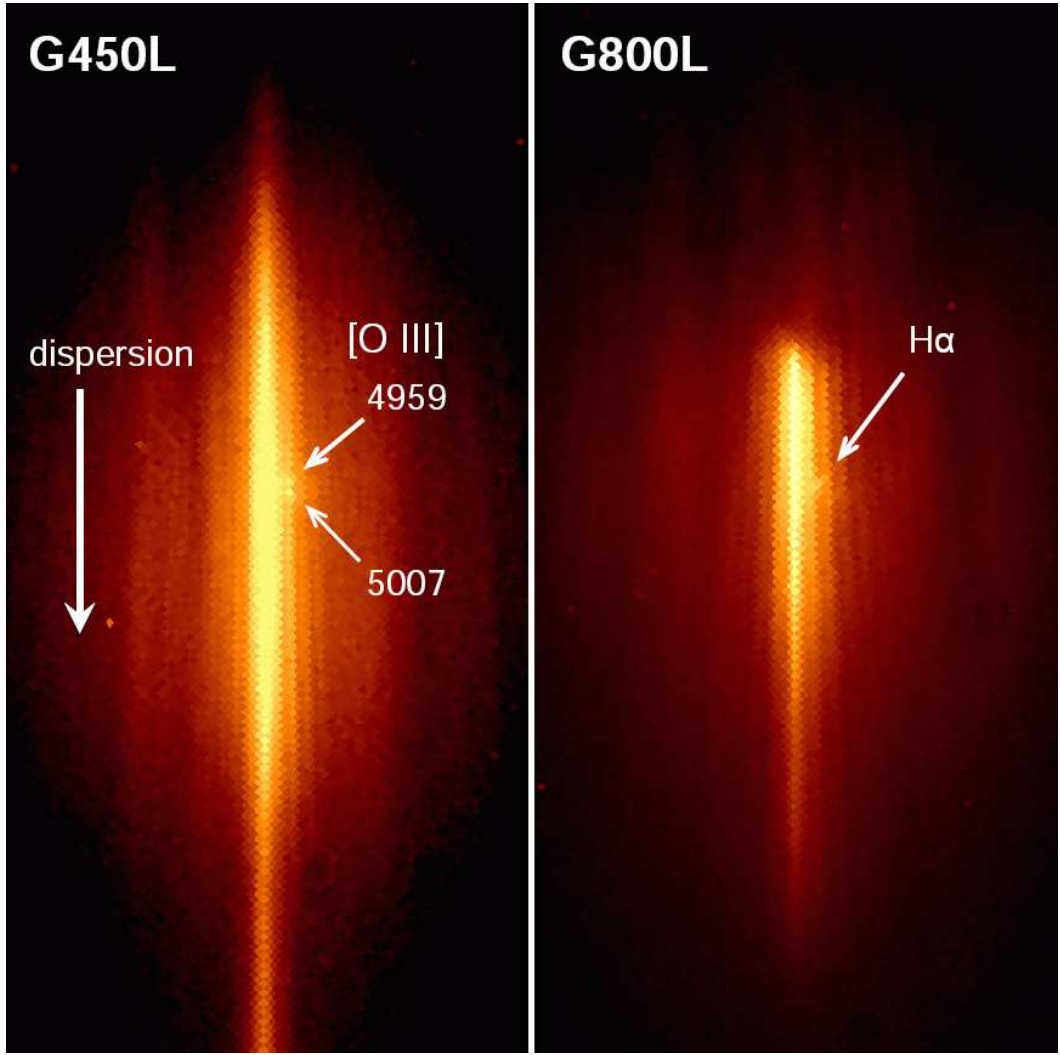}
\end{center} 
\caption{{\it HST\/} grism images, obtained with WF/PC1 in 1993. Both
images were taken at a telescope roll angle that maximized the spatial
separation of the faint companion from the hot white dwarf. 
{\bf Left:} G450L image, showing the intense blue spectrum of the hot
central star. The companion is detected only at the [\ion{O}{3}] 4959
and 5007~\AA\ nebular emission lines. {\bf Right:} G800L image,
showing the red continuum of the white dwarf. The companion is
detected only at the H$\alpha$ emission line. See text for further
discussion.}
\end{figure}

\begin{figure} 
\begin{center}
\includegraphics[width=5.25in]{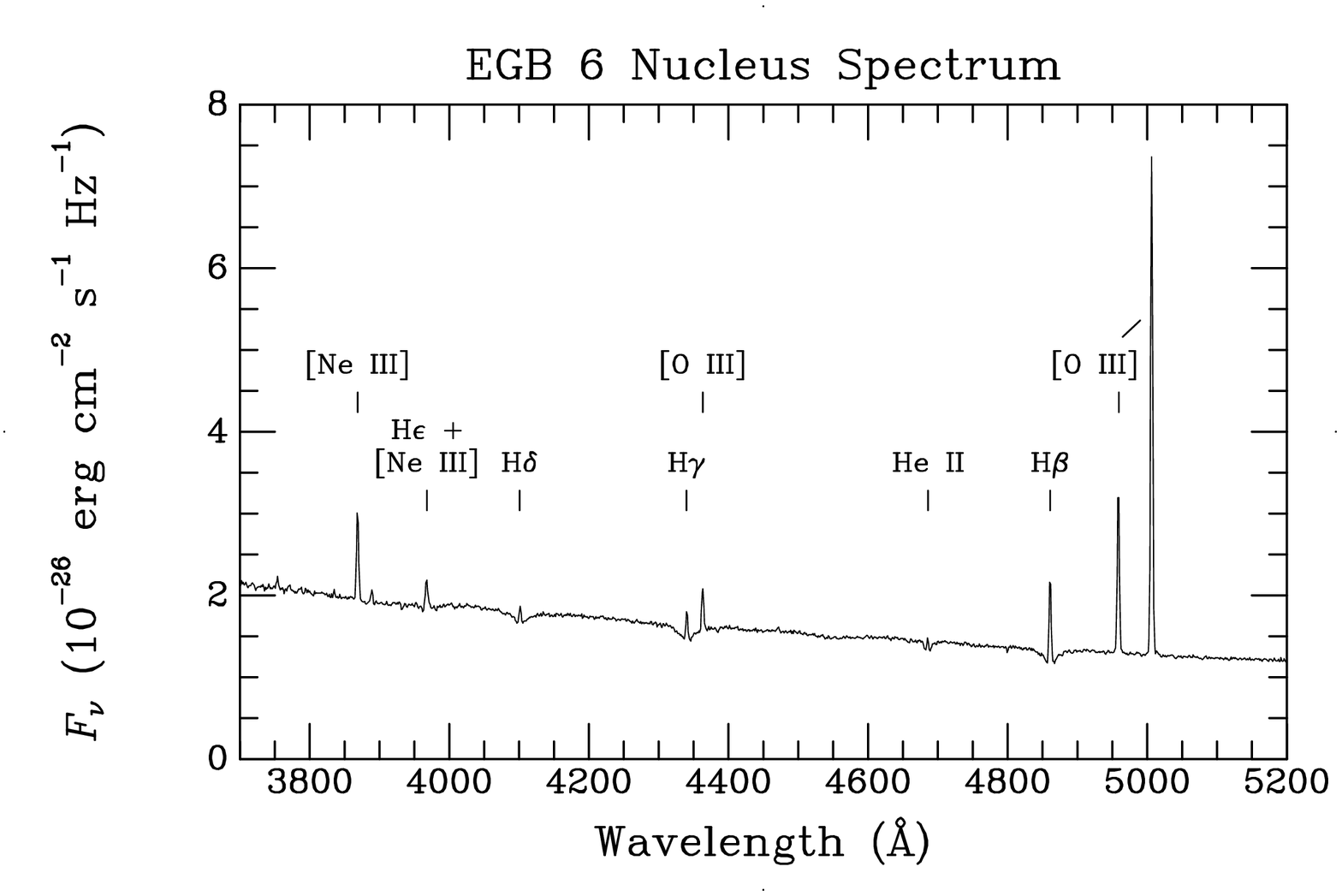}
\includegraphics[width=5.25in]{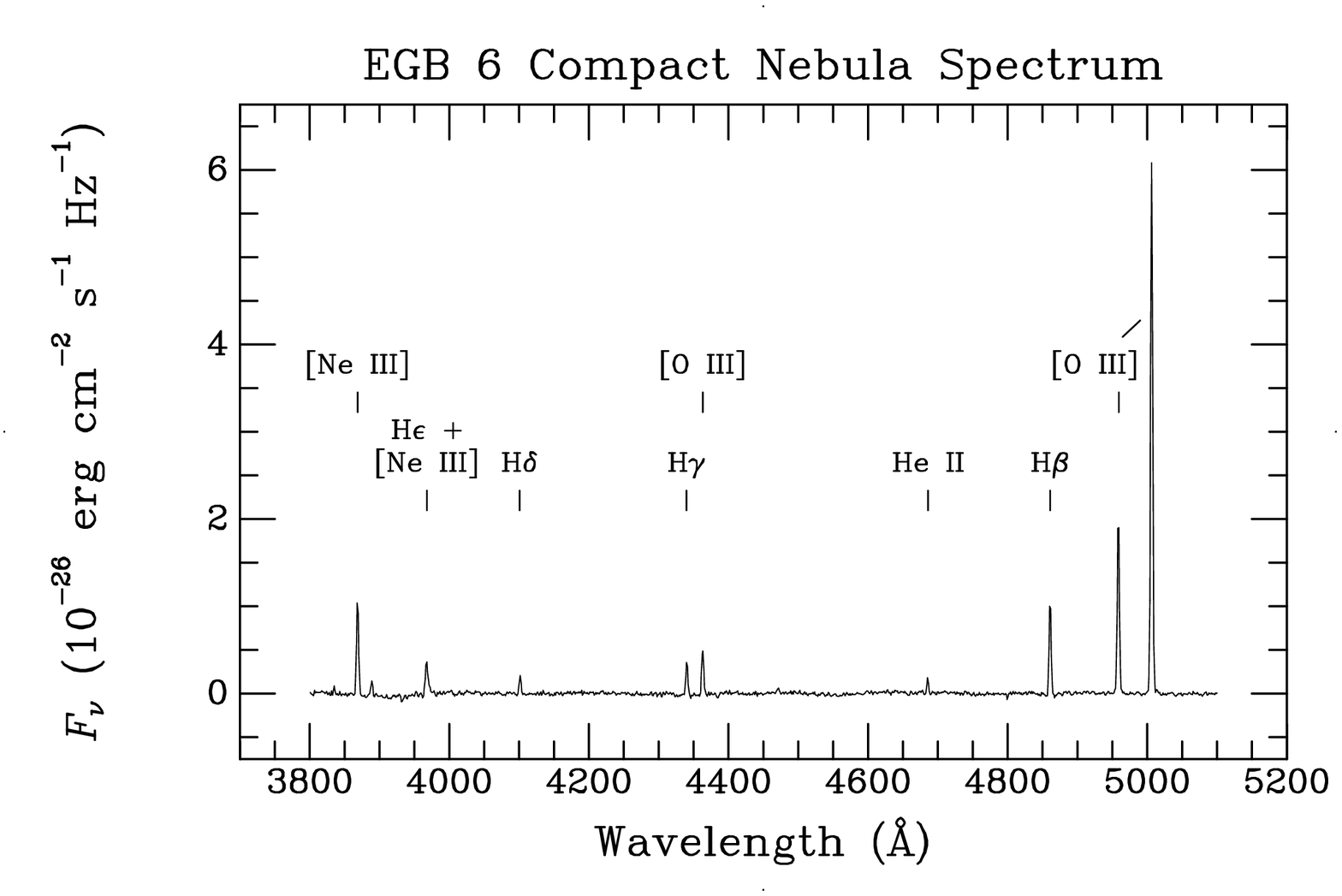}
\end{center} 
\caption{{\bf (top)} Spectrum of the central star of EGB~6 from the 2007 
6.5-m MMT observation, showing the photospheric spectrum of the DAOZ 
nucleus with superposed CEN\null. 
{\bf (bottom)} Emission-line spectrum of the CEN, obtained by
subtracting the best-fitting model-atmosphere WD synthetic
spectrum (Gianninas et al.\ 2010, G10), as described in the text.}
\end{figure}

\begin{figure} 
\begin{center}
\includegraphics[width=5.25in]{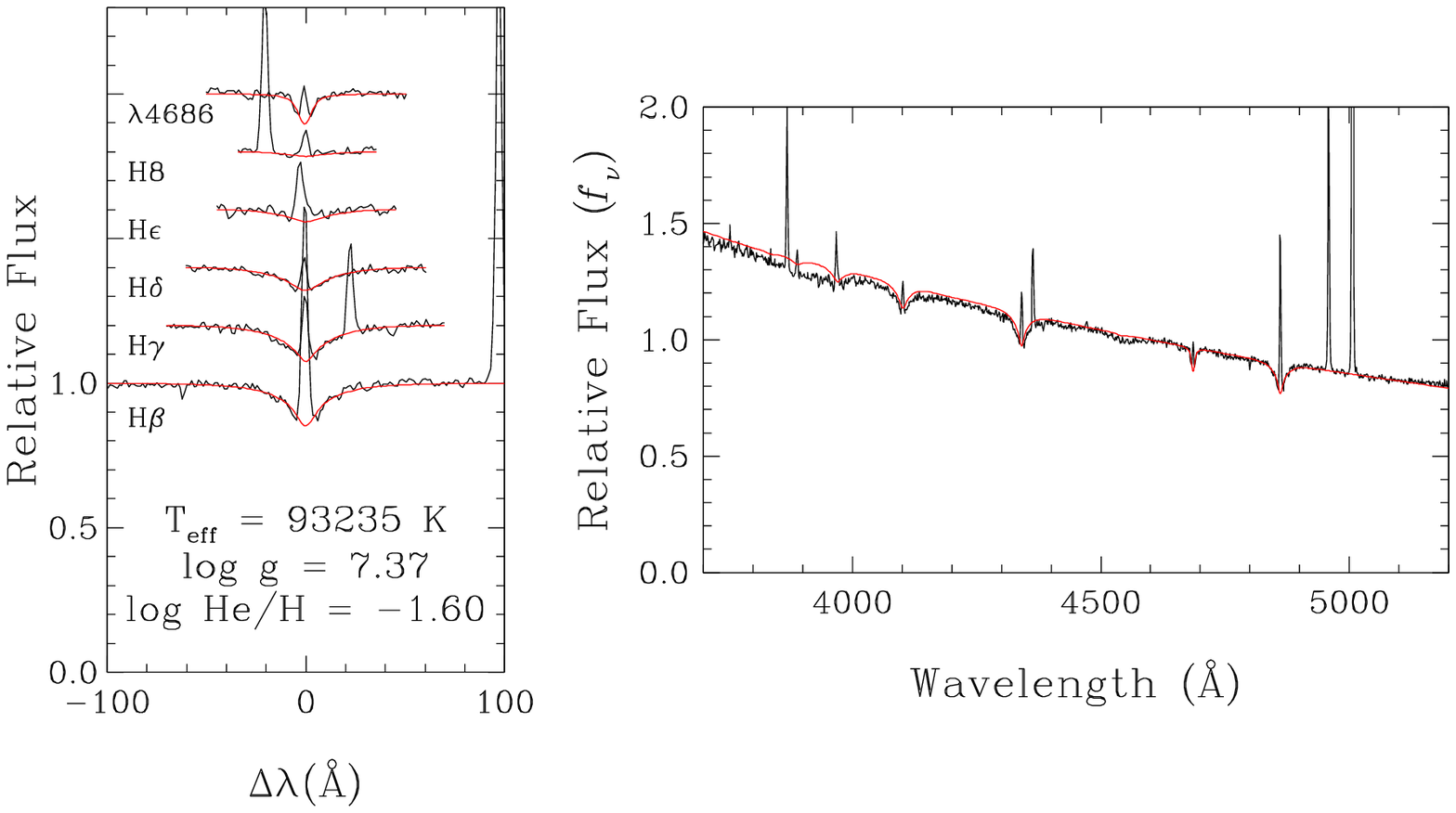}
\end{center} 
\caption{{\bf (Left)} Detailed fits to the He~II and individual Balmer 
absorption lines with the blended CEN emission lines omitted from the 
fitting procedure. The synthetic spectrum fit to the lines is shown as 
the red curve. 
{\bf (Right)} Superposition of the spectroscopic solution (red) to 
entire blue spectrum, both normalized to a continuum set to unity at
4600~\AA. }

\end{figure}


\clearpage

\begin{references}

\reference{} Bil\'ikov\'a, J., Chu, Y.-H., Gruendl, R.~A., Su,
K.Y.L., \& De Marco, O. 2012, \apjs, 200, 3 

\reference{} Bond, H.E. 1993, in {\it White Dwarfs: Advances in 
Observation and Theory}, NATO ASI, Series C, ed. M.A. Barstow, 
(Dordrecht: Kluwer) vol. 403, p. 85. 

\reference{} Bond, H.E. 2009, {\it J. Phys.: Conf. Ser.} vol. 172,
Number 1, 012029 

\reference{} Bond, H.E., Ciardullo, R., \& Meakes, M.~G. 1993,
{\it Planetary Nebulae}, Proc. IAU Symp. 155, eds. R. Weinberger 
\& A. Acker, (Kluwer Acad. Publ.: Dordrecht), v. 155, p. 397

\reference{} Bond, H.E., Liebert, J.W., Renzini, A., \& Meakes,
M.G. 1992, in {\it Science with the Hubble Space Telescope}, Proc. 
of a ST-ECF/STScI Workshop, ESO Conference and Workshop
Proceedings, No. 44, Garching near Munich: European Southern 
Observatory, eds. P. Benvenuiti and T.J. Schreier, p. 139

\reference{} Bond, H.E., Meakes, M.G., Liebert, J.~W., \& Renzini,
A. 1993, {\it Planetary Nebulae}, Proc. IAU Symp. 155, eds. R. 
Weinberger \& A. Acker, (Kluwer Acad. Publ.: Dordrecht) p. 499. 

\reference{} Bond, H.~E., Pollacco, D.~L., \& Webbink, R.F. 2003,
\aj, 125, 260

\reference{} Boumis, P., Paleologou, E.V., Mavromatakis, F., \&
Papamastorakis, J. 2003, \mnras, 339, 735 

\reference{} Chu, Y.-H., et al. 2011, \aj, 142, 75

\reference{} Ciardullo, R., Bond, H.E., Sipior, M.S., et al.\ 1999,
\aj, 118, 488

\reference{} Coleman, G.D., \& Worden, S.P. 1976, \apj, 205, 475

\reference{} Dopita, M.A., \& Liebert, J. 1989, \apj, 347, 910 (DL) 

\reference{} Ellis, G.L., Grayson, E.T., \& Bond, H.E. 1984, \pasp,
96, 283 (EGB) 

\reference{} Ferguson, D.H., Liebert, J., Cutri, R., Green, R.F., 
Willner, S.P., Steiner, J.E., \& Tokarz, S. 1987, \apj, 316, 399

\reference{} Fleming, T.A., Liebert, J., \& Green, R.F.  1986, \apj,
308, 176

\reference{} Frew, D.J., \& Parker, Q.A.\ 2010, \pasa, 27, 129 

\reference{} Fulbright, M.S., \& Liebert, J. 1993, \apj, 410, 275

\reference{} Gianninas, A., Bergeron, P., Dupuis, J., \& Ruiz, M.T. 
2010, \apj, 720, 581 (G10) 

\reference{} Gianninas, A., Bergeron, P., \& Ruiz, M.T. 2011, \apj, 
743, 138

\reference{} Green, R.F., Schmidt, M., \& Liebert, J. 1986, \apjs, 61,
305

\reference{} Hawley, S.A. 1981, \pasp, 93,  93 

\reference{} Holberg, J.B., \& Bergeron, P. 2006, \aj, 132, 1221 

\reference{} Ireland, M.J. et al. 2007, \apj, 662, 651

\reference{} Jacoby, G.H., \& van de Steene, G.\ 1995, \aj, 110, 1285

\reference{} Jeffries, R.D., \& Stevens, I.R. 1996, \mnras, 279, 180 

\reference{} Liebert, J., Bergeron, P., \& Holberg, J.B. 2005, \apjs,
156, 47

\reference{} Liebert, J., Bond, H.E., Dufour, P., \& Ciardullo, R.
2013, in {\it 18th European White Dwarf Workshop}, August 13-17, 2012,
in Krakow, Poland, ed. J. Krzesi\'nski, ASP Conf. Ser., in press.

\reference{} Liebert, J., Green, R., Bond, H.E., et al.\ 1989, \apj,
346, 251

\reference{} Miszalski, B., Boffin, H.M.J., Frew, D.J., et
al. 2012, \mnras, 419, 39

\reference{} Miszalski, B., Miko\'oajewska, J., K\"oppen, J., Rauch, T.,
Acker, A., Cohen, M., Frew, D.J., Moffat, A.F.J., Parker, Q.A., Jones,
A.F., \& Udalski, A. 2011, \aap, 528, A39 

\reference{} Nota, A., Jedrzejewski, R., Voit, M., \& Hack, W. 1996,
FOC Instrument Handbook Version 7.0 (Baltimore: STScI)

\reference{} O'Dell, C.R., Balick, B., Hajian, A.R., Henney, W.J., \&
Burkert, A. 2002, \aj, 123, 3329

\reference{} Osterbrock, D.E. 1974, in {\it Astrophysics of Gaseous
  Nebulae}, (San Francisco: Freeman)

\reference{} Osterbrock, D.E. 1989, in {\it Astrophysics of Gaseous 
Nebulae and Active Galactic Nuclei}, (University Science Books: Mill 
Valley CA)

\reference{} Schlafly, E.F., \& Finkbeiner, D.P. 2011, \apj, 737, 103 

\reference{} Schlegel, D.J., Finkbeiner, D.P., \& Davis, M. 1998,
\apj, 500, 525 

\reference{} Seaton, M.J. 1979, \mnras, 187, 73p

\reference{} Siegel, M.H., Hoversten, E., Bond, H.E., Stark, M., \&
Breeveld, A.A.\ 2012, \aj, 144, 65

\reference{} Smith, M.G., Aguirre, C., \& Zemelman, M. 1976, \apjs,
32, 217

\reference{} Su, K., Bil\'ikov\'a, J., Chu, Y.-H., et al.\ 2011, in 
{\it Asymmetric Planetary Nebulae 5}, Jodrell Bank Centre for
Astrophysics, eds. A.A.  Zijlstra, F. Lykov, E. Lagadec, \&
J. McDonald, p. 1.


\reference{} Su, K.Y.L., Chu, Y.-H., Rieke, G.H., et al.\ 2007,
\apjl, 657, L41

\reference{} Tweedy, R.W., \& Kwitter, K.B.\ 1996, \apjs, 107, 255

\reference{} Veeder, G.J. 1974, \aj, 79, 1056

\reference{} Wachter, S., Hoard, D.W., Hansen, K.H., Wilcox, R.E.,
Taylor, H.M., \& Finkelstein, S.L. 2003, \apj, 586, 1356

\reference{} Wood, B.E., Linsky, J.L., M\"uller, H.-R., \& Zank, G.P. 
2001, \apjl, 547 L49

\reference{} Wood, B.E., M\"uller, H.-R., Zank, G.P., \& Linsky, J.L. 
2002, \apj, 574, 412

\reference{} Zacharias, N., Finch, C.T., Girard, T.M., Henden, A., 
Bartlett, J.L., Monet, D.G., \& Zacharias, M.I. 2012, {\it UCAC4 
Catalog}, VizieR On-line Data Catalog: I/322. (see also SIMBAD)  

\reference{} Zuckerman, B., Becklin, E.E., \& McLean, I.S.\ 1991,
in {\it Astrophysics with Infrared Arrays}, ASP Conf. Series, vol.  
14, ed. R. Elston, (San Francisco: ASP), p. 161

\end{references}
\end{document}